\documentstyle[12pt]{article}
\begin{document}
\title{Asymptotic states in long Josephson junctions in an external
       magnetic field}
\author{K.N. Yugay, N.V. Blinov, and I.V. Shirokov \\
\em {Omsk State University, Omsk Institute of Sensor
      Microelectronics} \\
\em {of Russian Academy of Sciences,Omsk 644077, Russia}}
\date{}
\maketitle
\begin{abstract}
     Asymptotic states in long Josephson junctions  are investigated
in an external magnetic field. We show that a choice one of the
solution of the stationary Ferrell-Prange equation
can carry be out with use of an asymptotic solution of the sine-Gordon
equation and that an
evolution to that stable solution occurs by passing through
metastable states, which is determined with a form of quickly damped
initial perturbation. The boundary sine-Gordon and
Ferrell-Prange problems were carried out with a numerical simulation.
An approximated expression for the vortex and antivortex states
is obtained in the case of large values of an external magnetic field.
\end{abstract}

   The problem of magnetic field penetration in long Josephson junctions is
in a sense a classical one \cite{r1}--\cite{r5}. Nevertheless,
interest in it has not abated \cite{r6}--\cite{r13}. For example,
in long Josephson junctions a
dynamical chaos was discovered \cite{r6}--\cite{r10}.
A long Josephson junction is a very
good system to study Josephson vortex (fluxons, solitons)
motions and vortex interactions \cite{r12}--\cite{r13}.

  It is well known that a weak magnetic field penetrates exponentially
in a long Josephson junction. In a strong magnetic field a state of the
long Josephson junction arises that is  characterized by the appearance of
a Josephson vortex. An equation that describes this phenomenon is nonlinear
and in a general case it is the driven damped sine-Gordon equation.
In a stationary case we have the nonlinear Ferrell-Prange equation and
one would think the stationary states of a long Josephson junction represent
solutions of the above equation, although it turned out that the
Ferrell-Prange equation has more than one solution and the number of them
increases by increasing  the total length $L$ of the
long Josephson junction. Consequently, the problem of the selection of
solutions arises. Considering this problem from the thermodynamic
point of view one can affirm that a solution is realized which corresponds
to a minimum of the thermodynamic potential. However, the situation turns out
to be more complicated. In fact, if we suppose that the minimum of the thermodynamic
potential is satisfied with some solutions of the stationary equation,
then this criterion for selection of the solutions turns out to be insufficient.
Apparently this is precisely what takes place in our
case. In the present work it is shown that the problem of selection of
solutions for the stationary Ferrell-Prange equation is unequivocally solved
in the following way:  An asymptotic solution of the nonstationary
sine-Gordon equation is found that coincides with the solution of the
stationary Ferrell-Prange equation by $t\rightarrow\infty$,
and what is more either of asimptotic solution, which is realized,
depends on a form of rapid damped initial perturbation or on a way of
switching on an external magnetic field. This exceptionally important role of
initial perturbation that appears as a trigger mechanism is obviously
explained by system nonlinearity.

  We consider a long Josephson junction in an external magnetic field
$H_{ext}$ perpendicular to the junction. The Josephson phase variable
$\varphi (x,t)$ in a long Josephson junction is satisfied with
the one-dimensional sine-Gordon equation
\begin{equation}
  \varphi_{tt}(x,t)+2\gamma\varphi_t(x,t)-
  \varphi_{xx}(x,t)=-\sin\varphi(x,t),
\end{equation}
where $x$ is the distance along the junction normalized to the Josephson
penetration length $\lambda_J$,
$$
     \lambda_J=\Bigl({c\Phi_0\over 8\pi^2 j_c d}\Bigr)^{1/2},
$$
$t$ is the time normalized  to the inverse of the
Josephson plasma frequency $\omega_J$,
$$
  \omega_J=\Bigl({2\pi c j_c\over C\Phi_0}\Bigr)^{1/2},
$$
$\gamma$ is the dissipative coefficient per unit area, $\Phi_0$
is the flux quantum,
$j_c$ is the ctitical current density of superconductor,
$d=2\lambda_L+b$,
$\lambda_L$ is the London penetration length,$b$ is the
thickness of the dielectric
barrier, $C$  is the junction capacitance per unit area.

  Conditions corresponding to the presence of the external magnetic field
can be modeled by the following boundary conditions:
\begin{equation}
  \varphi_{x}(x,t)\vert_{x=0}=\varphi_{x}(x,t)\vert_{x=L} =
   H_{ext}(0,t)=H_{ext}(L,t).
\end{equation}
Here and below the magnetic field is normalized to
$\Phi_0/2 \pi\lambda_J d$, and the total length
of the junction $L$ is normalized to $\lambda_J$.

First of all we consider an asimptotic solution
(by $t\to\infty $) of
Eq.~(1) with the boundary conditions (2) and
suppose that $\lim_{t\to\infty}H_{ext}(0,t)=H_0$.
At first sight this problem is trivial,
since all asymptotic solutions will be stationary
on account of dissipation, they will be satisfied with the stationary
Ferrell-Prange equation with the corresponding boundary conditions:
\begin{equation}\begin{array}{c}
  \varphi_{0xx}(x)=\sin \varphi_0(x),
  \varphi_{0x}(x)\vert_{x=0}=\varphi_{0x}(x) \vert_{x=L}=H_0.
\end{array}\end{equation}

  However, numerical simulations of the boundary problem (1),(2) and (3) show
that not all solutions of the problem (3) are asymptotic solutions of the
problem (1), (2) or in other words we can say that not all solutions of the
problem (3) are physically observed. For example, for the small
external fields $H_0<1$ there exists a solution of the problem (3) that
corresponds to an observed dumping of the magnetic field into a
junction. Simultaneously with that solution another solution
of the problem (3) exists that gives very strong growth of a
field in the same direction into a long Josephson junction.
To cut off ``nonphysical'' solutions supplementry
considerations are attracted in the form of a condition of the minimum of a
thermodynamic potential, as was mentioned above.

   The results of problem (1), (2) simulations showed that
``nonphysical'' solutions are never realized in the asymptotic
approximation. The reason for the lack of the asymptotic ``nonphysical'' solutions
is their instability relative to small perturbations.

We show that the stationary solution $\varphi_0(x)$ of the problem (3)
is stable then and only then when all values of the spectrum $E$
are positive, i.e., $E_{\rm min} > 0$, for the boundary problem
\begin{equation}\begin{array}{c}
  -u_{xx}(x) + u(x) \cos \varphi_0(x) = E u(x), \\[10pt]
   u_x(x)\vert_{x=0}= u_x(x)\vert_{x=L} = 0 .
\end{array}\end{equation}

    To prove this assertion we shall linearize the sine-Gordon
equation in the vicinity of stationary solution $\varphi_0(x)$;
i.e.,  we shall
suppose $\varphi(x,t)=\varphi_0(x)+\theta(x,t)$, where $\theta(x,t)$
is an infinitesimal perturbation.
We obtain the equation for the function $\theta(x,t)$ from the
sine-Gordon equation taking into account the Ferrell-Prange equation:
\begin{equation}\begin{array}{c}
  \theta_{tt}(x,t)+ 2\gamma \theta_t (x,t) -
  \theta_{xx}(x,t) = - \theta(x,t) \cos\varphi_0(x), \\[10pt]
  \theta_x(x)\vert_{x=0}= \theta_x(x)\vert_{x=L} = 0 .
\end{array}\end{equation}
We can obtain a general solution of the boundary Eq.~(5) by means of
the expansion of the function $\theta(x,t)$ in a series in terms of a complete
system of eigenfunctions of the Schr\"odin\-ger operator with the
potential $\cos\varphi_0(x)$:
\begin{equation}
   \theta(x,t) = \sum_n e^{\lambda_n t}u_n(x),
\end{equation}
where $\lambda_n$ and $u_n(x)$ are eigenvalues and eigenfunctions of
the  Schr\"odin\-ger operator of the problem (4), respectively. Substituting
expansion (6) into Eq.~(5) and taking into account Eq.~(4),
we get for $\lambda_n$:
\begin{equation}
     \lambda_n = -\gamma \pm \sqrt{\gamma^2 - E_n}.
\end{equation}
It is clear that if even one of the numbers $\lambda_n$
has a positive real part,
then the infinitesimal perturbations $\theta(x,t)$ will increase
exponentially
in the course of time and, further,
if all $\lambda_n$ have a negative real part, then
the perturbations damp exponentially. Since the Schr\"odinger operator
is a Hermitian operator in a space of functions which  have a derivative
on the ends of intervals equal to zero, its eigenvalues $E_n$ are real.
It results
from expression (7) that $\lambda_n$ can have a positive real part only
with $E_n < 0$. Thus with $E_n > 0$ all $\lambda_n <  0$ and
perturbations $\theta(x,t)$ damp exponentially.
It is clear that the condition $E_{n \rm min} > 0$ is the condition
of the stable solution of the problem (3) with any values $\gamma > 0$.

   One can call unstable solutions $\varphi_0(x)$ of
the problem (3) metastable,
since they may be long-lived. Analysis of the spectrum problem gives
information about the survival time of metastable states.
Let $\Re\lambda_n>0$;
then $\tau = (\Re \lambda_n)^{-1}$ makes sense of the characteristic
decay time of the metastable state.
The corresponding eigenfunctions $u_n(x)$ are
perturbations that destroy the metastable states.

A detailed analysis of the problem (1), (2), (3), and (4) is failed
to be carried out by analytical methods. The numerical simulation
results of this problems showed the following.

    (1) The number $N$ of stable and metastable states, i.e., the number of
solutions of the problem (3), increases with increasing $L$.
For example, with $H_0 > 2$ for $L = 2 \pi$, $N = 4$, for $L = 20$, $N = 8$.

    (2) From all solutions of the problem (3) approximately half
of the states is stable.
For example, with $H_0=3$ for $L = 2 \pi$, 2 states out of
4 are stable; for $L=20$, 3 states out of 8 are stable;
for $L = 40$, 11 states of 18 are stable.

    (3) The numerical investigations of the sine-Gordon Eq.~(1)
showed that either stable state [stable solution of the
problem (3)] is realized depending on the method of superposition
of an external field in a long Josephson junction. In other
words, an established magnetic field and currents in the junction
``remember'' how an external magnetic field was started.

  For the case $H_0 = 2$, $L = 2 \pi$  switching on the magnetic
field in the form
$H_{\rm ext}(0,t)=H_{\rm ext}(L,t)
=H_0[1-a \exp(-t/5)\cos t]$ leads to
the system coming to one or another state (Fig.~1) depending on the
controlling parameter $a$. So with $0\le a\le 0.36$ state 4
is realized in Fig.~1;
with $0.37\le a\le 0.74$ the one fluxon state 6 is realized;
with $a\ge 0.75$ the two fluxon state 8 is realized.

 It turns out that transition to one of the stable asymptotic
states occurs through the metastable states. For example, with $a=0.74$
state 6 arises from the metastable state 7; with $a=0.75$
state 8 arises from the same metastable state 7 (Fig.~1). This
property of the system is not characteristic for the case $H_0 = 2$,
corresponding to a separatrix. For other values of $H_0$
the same property takes place.
In Fig.~2 the evolution of the magnetic field is shown for the case $H_0=2$
with different values of the parameter $a$ into the junction.

  In our opinion the conclusion is the most important. Thus,
an asymptotic state of a magnetic field, realized in a long Josephson
junction depends essentially on a method of switching on
an external field, which is a direct result of the sine-Gordon equation.
The asymptote of problem (1), (2) taking into account the
method of switching on an external field does select one of the stable
solutions of the stationary Ferrell-Prange equation. It is
possible that this property of the solution is general enough
for nonlinear systems. With the condition of minimum thermodynamic potential
all stable solutions of problem (3) are satisfied apparently.

We should like to note another interesting fact following from the numerical
simulation of problems (1), (2) and (3). Among all the stationary
solutions there exists one which corresponds to the maximum of
a magnetic field and the another that corresponds to the minimum
of a magnetic field. In Fig.~1 these are states 8 and 6,
respectively.
The states are stable. State 8 in this figure is called a vortex;
state 6 in the same figure is called an antivortex. For these states we can
obtain an approximate formula solving problem (3) by means of the
iteraction method
\begin{equation}\begin{array}{c}
  \varphi_0(x) \equiv \lim_{k \to \infty} \varphi_0^{(k)}(x), \\[10pt]
  \varphi^{(k)}_{0xx}(x) = \sin \varphi_0^{(k-1)}(x) , \\[10pt]
  \varphi^{(k)}_{0x}\vert_{x=0}= \varphi^{(k)}_{0x}\vert_{x=L} = H_0 .
\end{array}\end{equation}
Suppose $\varphi_0^{(0)}(x) = H_0 x + \alpha$,
where $\alpha$ is some kind of constant. Then
\begin{equation}\begin{array}{c}
  \varphi^{(1)}_{0xx}(x) = \sin( H_0 x+ \alpha),\\[10pt]
  \varphi^{(1)}_{0x}\vert_{x=0}= \varphi^{(1)}_{0x}\vert_{x=L} = H_0 .
\end{array}\end{equation}
Integrating (8), we find
\begin{equation}
H^{(1)} \equiv \varphi_{0x}^{(1)}(x) = H_0 \pm
{1\over H_0} \Bigl[ \cos\bigl(H_0 x - {H_0 L \over 2}\bigr)  -
                     \cos\bigl({H_0 L \over 2}\bigr) \Bigr] .
\end{equation}
In formula (10) the ``$+$'' sign corresponds to a vortex state;
the ``$-$'' sign corresponds to an antivortex state. Expression (10)
is correct in the case of large fields. Comparision of the
values, obtained by means of formula (10), with results of
numerical simulation shows that they practically do not
differ for $H_0=10$.

In conclusion we note that a swing of deflections of the field $H^{(1)}(x)$
from $H_0$ makes $\approx 2/H_0$ and this swing decrease by increasing
of the external field. In other words one can say that the number of the
stationary states per unit of field increases. In turn this
circumstance may lead to the small fluctuations of the external
field $H_0$  to cause of a ``jump'' from the one stable state to the
other. We hope to discuss this problem in detail in our next
work.

\newpage
\thispagestyle{empty}
\begin{enumerate}
\item[Fig. 1.] Solutions of boundary problem (3)
  for the case $H_0=2, L=2\pi$. Solid line (4,6,8), stable
  states; dashed line (1,2,3,5,7), metastable states.

\item[Fig. 2.] Evolution of magnetic field
  for the case $H_0=2, L=2\pi$ with different
  values of parameter $a$.
  (a) $a=0.2$, (b) $a=0.5$, and (c) $a=1.0$.

\end{enumerate}
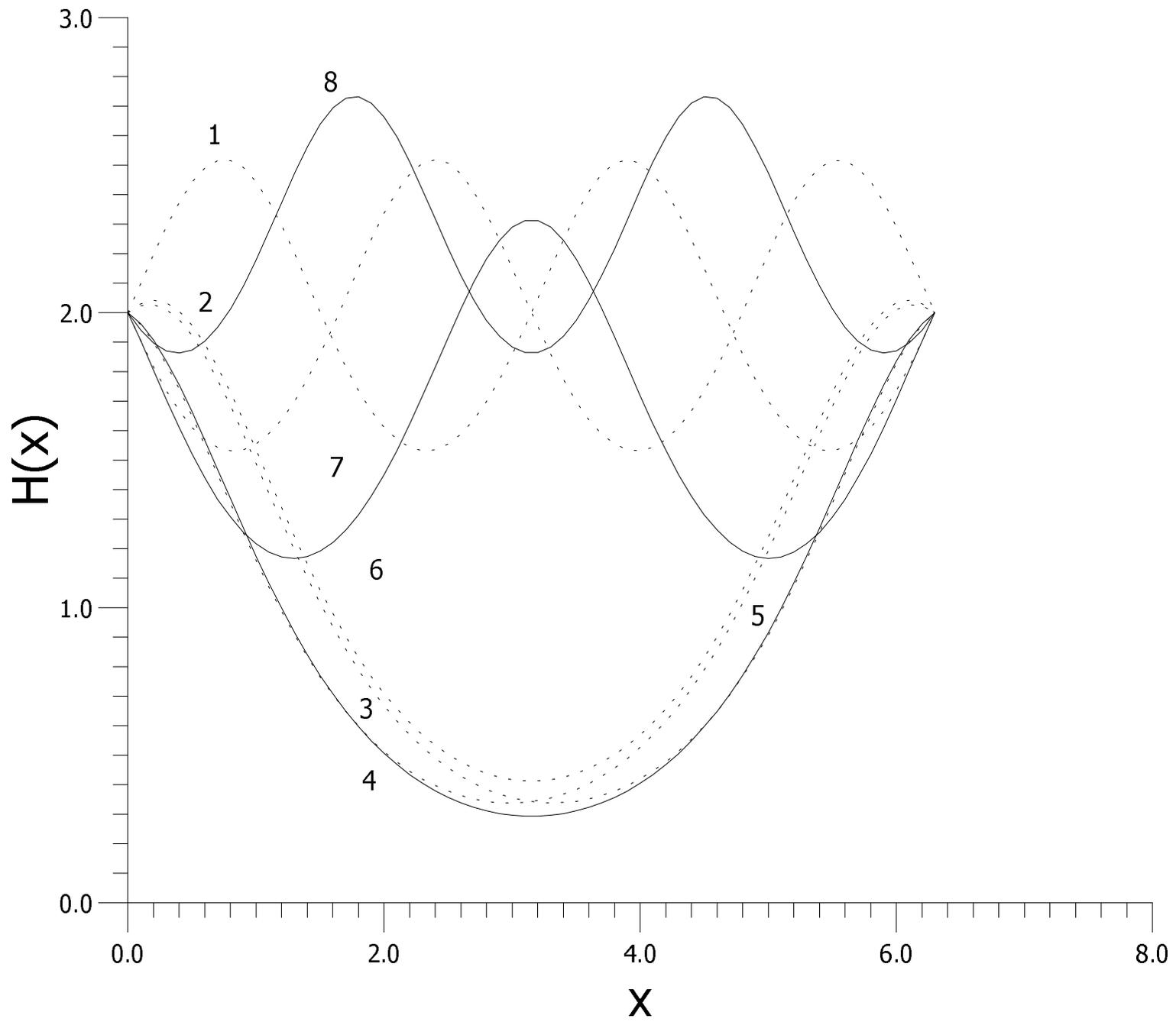
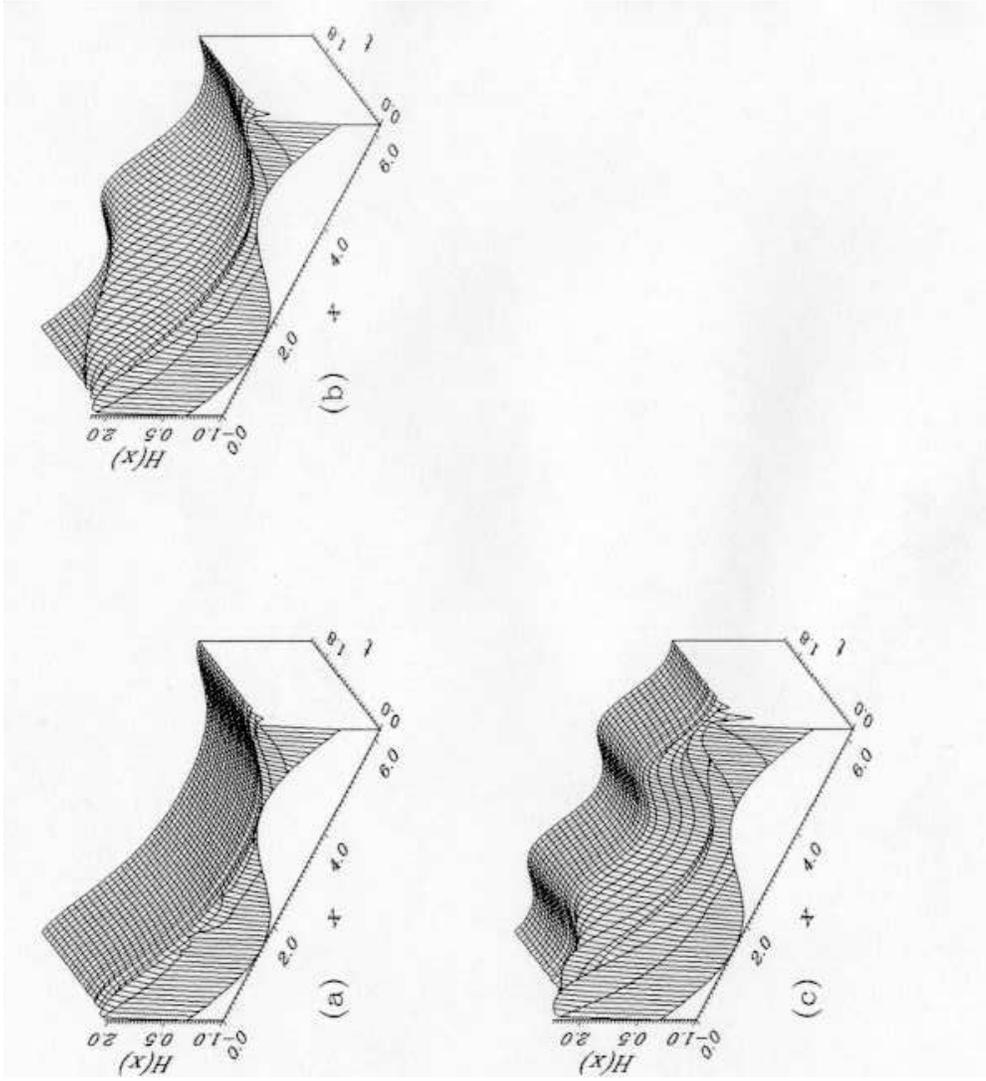
\end{document}